\documentclass[12pt,a4paper]{article}
\usepackage{amsmath,amssymb,graphicx,color}
\begin{document}
\newcommand{\be}{\begin{equation}}
\newcommand{\bl}[1]{\begin{equation}\label{#1}}
\newcommand{\ee}{\end{equation}}
\newcommand{\pd}[2]{\frac{\partial{#1}}{\partial{#2}}}
\newcommand{\z}[1]{\left({#1}\right)}
\newcommand{\sz}[1]{\left[{#1}\right]}
\newcommand{\kz}[1]{\left\{{#1}\right\}}
\renewcommand{\ae}[1]{\left|{#1}\right|}
\newcommand{\rec}[1]{\frac{1}{#1}}
\newcommand{\m}[1]{\mathrm{#1}}
\renewcommand{\c}[1]{\mathcal{#1}}
\renewcommand{\v}[1]{\mathbf{#1}}
\renewcommand{\r}[1]{(\ref{#1})}

\textwidth=135mm
 \textheight=200mm
\begin{center}
{\bfseries 
Oscillating HBT radii and the time evolution of the source  -- 
$\sqrt{s_{NN}} = 200$ GeV Au+Au data analyzed with azimuthally sensitive Buda-Lund hydro model
\footnote{{\small Presented by T. Cs\"org\H{o} at the Workshop on Particle Correlations and Femtoscopy, Kiev, Ukraine, September 17, 2010.}}}
\vskip 5mm
   A.~Ster$^{1,2}$,
   M.~Csan\'ad$^{3}$,
   T.~Cs\"org\H{o}$^{1,4}$,
   B.~L\"orstad$^{2}$
   and B.~Tom\'a{s}ik$^{5,6}$
\vskip 5mm
{\small {\it $^1$ MTA KFKI RMKI, H-1525 Budapest 114, P.O. Box 49, Hungary\\
$^2$Department of High Energy Physics, University of Lund, S-22362 Lund, Sweden\\
$^3$E\"otv\"os University, H-1117 Budapest XI, P\'azm\'any P\'eter s. 1/A, Hungary\\ 
$^4$Department of Physics, Harvard University, 17 Oxford St, Cambridge, MA 02138, USA\\
$^5$Univerzita Mateja Bela, Tajovsk\'eho 40, SK-97401 Bansk\'a Bystrica, Slovakia\\ 
$^6$FNSPE, Czech Technical University in Prague, B{r}ehov\'a 7, CZ-11519 Prague, Czech Republic
}} 
\\
\end{center}
\vskip 5mm
\centerline{\bf Abstract}
Identified particle spectra of pions, kaons and (anti)protons, and elliptic flow and azimuthal
dependence of Bose-Einstein or HBT correlations of identified pions
in $\sqrt{s_{NN}} = 200$ GeV Au+Au  collisions is analyzed simultaneously using 
an ellipsoidally symmetric generalization of the  Buda-Lund hydrodynamical model.
The transverse flow is found to be faster in the reaction plane than out of plane, which
results in a reaction zone that gets slightly more elongated in-plane than out of plane.

\vskip 10mm
\section{Introduction}
\label{intro}

Important information about the properties of extremely hot strongly 
interacting matter comes from the observation of azimuthal anisotropies
in non-central ultra-relati\-vistic nuclear collisions. The second order Fourier  
component of azimuthal hadron distributions is connected with the azimuthal 
dependence of transverse 
collective expansion velocity of the bulk matter \cite{Heiselberg:1998es,Sorge:1998mk}.
That is in turn determined by the differences of the  initial pressure gradients
in the two perpendicular transverse directions, 
as well as by the initial geometry, 
the initial velocity and temperature distributions of the
fireball, and the equation of state~\cite{Heinz:2001xi,Csorgo:2001xm}. The anisotropic shape of the fireball 
measured with the help of correlation femtoscopy \cite{Wiedemann:1997cr} at the 
instant of final decoupling of hadrons bears information about the total lifespan 
of the hot matter: with time the originally out-of-reaction-plane shape becomes 
more and more round and may even become in-plane extended \cite{Heinz:2002sq}. Unfortunately, 
in determining the elliptic flow and azimuthally sensitive correlation radii 
individually two effects---spatial and flow anisotropy---are entangled. For example, 
the same elliptic flow can be generated with varying flow anisotropy strength
if the spatial anisotropy is adjusted appropriately~\cite{Tomasik:2004bn}.

In general, the precise way of the interplay 
between the two anisotropies is model dependent. It has been studied and shown 
to be different within the Buda-Lund model \cite{Csanad:2008af} than in 
   the Blast Wave model \cite{Tomasik:2004bn}. 

Here we report the results of the Buda-Lund hydro model analyzis 
of azimuthally sensitive Hanbury Brown -- Twiss (HBT) 
radii, using data from non-central heavy ion collisions at RHIC.
Note that the original, axially symmetric version of the Buda-Lund model described 
successfully data from central Au+Au collisions at RHIC,
as measured by BRAHMS, PHE\-NIX, PHOBOS, and STAR collaborations, 
including identified particle spectra and transverse mass dependent 
HBT radii as well as the pseudorapidity distributions of charged particles as
first presented in refs.~\cite{Csorgo:2002ry,Csanad:2003sz}.

The Buda-Lund model formalism for non-central collisions, including
elliptic flow and azimuthal angle dependence of HBT radii has been 
proposed first in ~\cite{Csanad:2008af}. The model is defined  with the help of 
its emission function. In order to take into account the effects of resonance decays
it uses the core-halo model \cite{Csorgo:1994in}. 
This ellipsoidal extension of the Buda-Lund model was shown before 
to describe well the transverse mass and the pseudorapidity
dependence of elliptic flow parameter $v_2$ of identified particles at various energies and
centralities in ref.~\cite{Csanad:2005gv}.

In the present study, we improve
on earlier versions of the Buda-Lund model, by scrutinizing the various components
using azimuthally sensitive HBT data. 
Eventually we utilize a model that includes as a special case of T.S. Bir\'o's  
axially symmetric and accelerationless exact solution of 
relativistic hydrodynamics~\cite{Biro:1999eh},
in contrast to the original, earlier variant, ref. \cite{Csanad:2008af}, which was based on
an ellipsoidally symmetric, but also non-accelerating exact solution of
relativistic hydrodynamics, given by ref. \cite{Csorgo:2003rt}.  
Similarly to ref. ~\cite{Csorgo:1999sj},
we utilize here an improved calculation, using the binary source formalism,
to obtain the observables by using two saddle-points instead of only one. This results in
an oscillating pre-factor in front of the Gaussian in the two-pion correlation function
that we take into account for the formulae of the HBT radii. Details of the model
and the evaluation of the observables from it are presented in ref.~\cite{Ster:2010ia}.

Azimuthally sensitive HBT radii 
were also considered recently in cascade models, e.g. in the fast Monte-Carlo
model of ref.~\cite{Amelin:2007ic}, or, in the Hadronic Resonance Cascade~\cite{Humanic:2005ye}.

Data analysis of correlation HBT radii performed earlier with the Blast Wave model 
indicates that the fireball at the freeze-out is elongated slightly out of the reaction 
plane \cite{Adams:2004yc}, i.e. spatial deformation is similar as in the initial state given by the 
overlap function. This is also supported by the theoretical results from 
hydrodynamic simulations \cite{Heinz:2002sq,Frodermann:2007ab} and URQMD \cite{Lisa:2009wp}. 
It sets limitations on the total lifespan. From all previous analyses it seems, 
however, that the final state 
anisotropy has an interesting non-monotonous dependence on collision energy with a minimum 
at the SPS energies \cite{Lisa:2009wp}. In our analysis of the \emph{same} data with a 
\emph{different} model we observe for the first time at RHIC an in-plane elongation 
of the fireball at freeze-out.

This presentation is a conference contribution which is based  
on a more a detailed manuscript~\cite{Ster:2010ia}, 
where we described the azimuthally sensitive Buda-Lund model fully 
and detailed the analytic formulas that 
were obtained from it and were fitted simultaneously to the single particle spectra 
of identified particles, the  elliptic flow and the azimuthal angle dependent HBT radii 
of identified pions.

%%%%%%%%%%%%%%%%%%%%%%%%%%%%%%%%%%%%%%%%%%%%%%%%%%%%%%%%%%%
\section{asBuda-Lund hydro compared to data}
\label{obs}
\label{fits}
Observables like spectra, elliptic flow or Bose-Einstein
correlation functions have been calculated analytically from the
azimuthally sensitive Buda-Lund hydrodynamic model, (or asBuda-Lund hydro in
a shortened form), using a double saddle-point approximation in the integration,
as detailed in refs.~\cite{Ster:2010ia,Csanad:2003qa}. 
We have determined~\cite{Ster:2010ia}  the best values of the model 
parameters by fitting these analytic, parametrically given 
expressions for the observables to experimental data
with the help of the CERN Minuit fitting package. 
This is possible given that the Buda-Lund hydro model relies on exact but
parametric solutions of (relativistic) hydrodynamics in certain limiting cases
when these solutions are known, and interpolates between these solutions in
other cases. So dynamics is mapped to time evolution of model parameters, and
hadronic observables are sensitive only to the values of these model parameters
around the time of freeze-out, as was explicitely demonstrated recently in ref.
~\cite{Csanad:2009wc}. However, penetrating probes, for example the direct
photon spectra, are known to carry explicit information about the equation of state
through the cooling history of the continuous radiation of these penetrating probes
for example direct photons. This property of the model was used recently
to determine the equation of state parameters of the strongly interacting 
quark-gluon plasma in 200 GeV Au+Au collisions in refs.~\cite{Csanad:2011jq,Csanad:2011jr}

Data from 20-30\% centrality class of 200 AGeV Au+Au collisions provided by 
PHENIX~\cite{Adler:2003cb,Adler:2004rq} 
and STAR~\cite{Adams:2004bi,Adams:2003ra} were used in this analysis.
The fits were performed simultaneously to azimuthally integrated 
transverse mass spectra of positive and negative pions, kaons, and 
(anti)protons \cite{Adler:2003cb}, the transverse momentum
dependence of the elliptic flow parameter $v_2$ of pions~\cite{Adams:2004bi}
and to the HBT radii due to pion correlations
as functions of transverse mass and the azimuthal angle~\cite{Adams:2003ra}. 
The results are plotted in Figs. 1-3. 

The interpretation of the model parameters is summarized in Table~\ref{t:pars}. 
The two radii $R_{sx}$ and $R_{sy}$ correspond to ``thermal surface" sizes,
corresponding to distances where the temperature drops to half of its central value, to
$T_0/2$, while parameter $T_e$ corresponds to the temperature of the center after
most of the particle emission is over (cooling due to evaporation and expansion). 
Sudden emission corresponds to the  $T_e = T_0$, and $\Delta\tau \rightarrow 0$ limit. 
Also note that we use $\mu_B$, baryochemical potential, 
calculated from the chemical potential of protons and antiprotons as
$\mu_B = 1/2 \  (\mu_{0,p}-\mu_{0,\overline{p}})$, see Table~\ref{t:pars}.
The flow profile is linear in both transverse directions but the Hubble constant
is direction dependent, denoted by $H_x$ and $H_y$ in the reaction plane, 
and out of the reaction plane, respectively.

In Table~\ref{t:results}, we present the model parameters obtained from simultaneous 
fits to the data sets. For comparison, results are shown from our earlier 
analysis of 0-30\% centrality collisions~\cite{Csanad:2004mm}, too, 
that was performed with a previous version of the model corresponding to the 
axially symmetric limit of the current ellipsoidally generalized 
Buda-Lund hydrodynamic model.

The general observation is that the Buda-Lund model parameters 
describing the source of non-central reactions
are usually slightly smaller than those of more central
collisions. However, the changes are usually within 2 standard deviations,
therefore the above statement is based on the tendency of the parameters,
and on some lower energy results not shown here but presented in 
ref.~\cite{Csanad:2004mm}, too.
For example, the central temperature in these particular
non-central reactions is below that of the more central ones. Also,
the transverse geometrical radii at the mean emission time are
considerably smaller compared to the more central values. 
Moreover, the geometric shape evolution due to 
the asymmetric particle transverse flow in-plane (x) an out-of-plane (y)
directions results in a source more elongated in-plane. Due to the
smaller longitudinal source size, the 
parameter corresponding to the formation of hydrodynamic phase is 
about 10\% \ smaller than that in more central collisions, 
$\tau_0 (20-30\%) = 5.4 \pm 0.1$~fm/$c$. The elongation in longitudinal
direction is similarly smaller, $\Delta\eta (20-30\%)= 2.5 \pm 0.3$. 
In both cases the baryochemical potential is found to be small 
as compared to the proton mass. We emphasize again that
the observations are based on all the fit results in 
ref.~\cite{Csanad:2004mm}. 

Note that some of the azimuthally sensitive data have large
systematic errors that affect the success of fits which we had to take 
into account. The reason for that is the difficulty of precise 
determination of the event reaction plane the data are relative to. 
Several methods are used by the experiments to overcome it and we 
mention those applied for the selected data. 

The data set we used for fitting $v_2$ was calculated by the 
four-particle cumulants reaction plane determination method that is based on 
calculations of $N$-particle correlations and non-flow effects
subtracted to first order when $N$ is greater than 2. The higher $N$ is the
more precise the event plane determination is, as expected.
STAR published two-particle cumulants $v_2$ data in the same reference, too,
but because of the visible deviations between the two kinds of data sets and 
with respect to the comments above we used $v_2\{4\}$ data, only. 
For further details, see ref.~\cite{Adams:2004bi}.

In case of azimuthally sensitive correlation radii, STAR has cast 
about 10\% possible systematic errors on the data on average. The most
likely deviations were assumed to take effect on the 'side' and 'out' radii
of transverse momentum of 0.2 GeV/$c$. The $\chi^2$/NDF for the full fit,
including HBT radii with their statistical errors is 269.6/152, 
which corresponds to a very low confidence level. 
But, when we tested our fits with the above mentioned two radii of 'side' and 'out' 
of transverse momentum of 0.2 GeV/$c$ shifting them within their 
systematic errors (about $\pm$ 5\%) we could achieve an
acceptable 1\% confidence level for the full simultaneous fit.
Without the contribution of the HBT radii to $\chi^2$/NDF 
the confidence level is of an acceptable level of 5.1\%. 

Our results were compared also to results of refs.~\cite{Retiere:2003kf,Tomasik:2004bn} of the 
azimuthally sensitive extension of the Blast Wave model in the detailed write-up, 
ref.~\cite{Ster:2010ia}.

%%%%%%%%%%%%%%%%%%%%%%%%%%%%%%%%%%%%%%%%%%%%%%%%%%%%%%%%%%%%%%%%%%%%%%%%%%%%%%%%%%%%%%%
\begin{figure}
	\centering
		\includegraphics[width=0.9\textwidth]{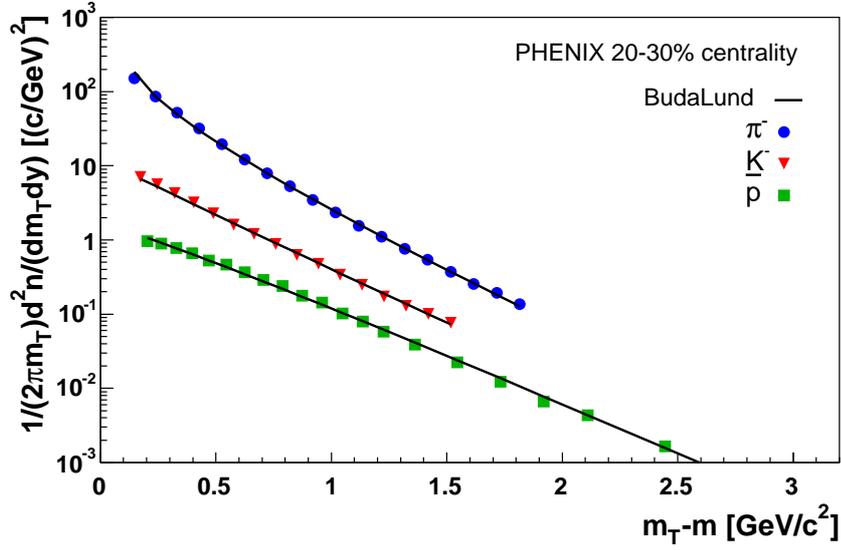}\\
		\includegraphics[width=0.9\textwidth]{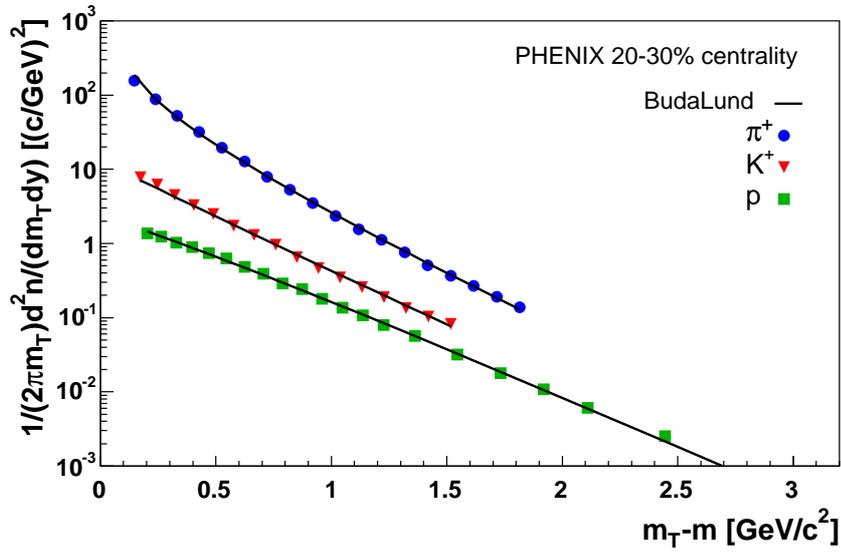}
\caption{Buda-Lund model fits to $\sqrt{s_{NN}}= 200$ GeV Au+Au data of ref.~\cite{Adler:2003cb}
on azimuthally integrated transverse momentum spectra of 
negatively (top figure)  and  positively (bottom plot) charged particles.
}
	\label{f:fig1}
\end{figure}
%%%%%%%%%%%%%%%%%%%%%%%%%%%%%%%%%%%%%%%%%%%%%%%%%%%%%%%%%%%%%%%%%%%%%%%%%%%%%%%%%%%%%%%
%%%%%%%%%%%%%%%%%%%%%%%%%%%%%%%%%%%%%%%%%%%%%%%%%%%%%%%%%%%%%%%%%%%%%%%%%%%%%%%%%%%%%%%%
\begin{figure}
	\centering
		\includegraphics[width=0.90\textwidth]{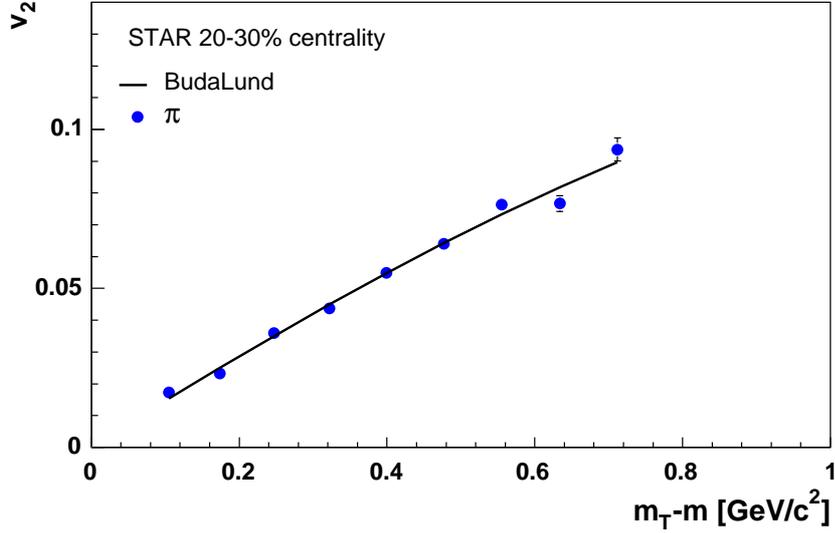}
\caption{Buda-Lund model fit to RHIC 200GeV Au+Au data
on $v_2$ elliptic flow of pions, data from ref.~\cite{Adams:2004bi}. }
\label{f:fig5}
\end{figure}
%%%%%%%%%%%%%%%%%%%%%%%%%%%%%%%%%%%%%%%%%%%%%%%%%%%%%%%%%%%%%%%%%%%%%%%%%%%%%%%%%%%%%%%
%%%%%%%%%%%%%%%%%%%%%%%%%%%%%%%%%%%%%%%%%%%%%%%%%%%%%%%%%%%%%%%%%%%%%%%%%%%%%%%%%%%%%%%
\begin{figure}
	\centering
		\includegraphics[width=0.90\textwidth]{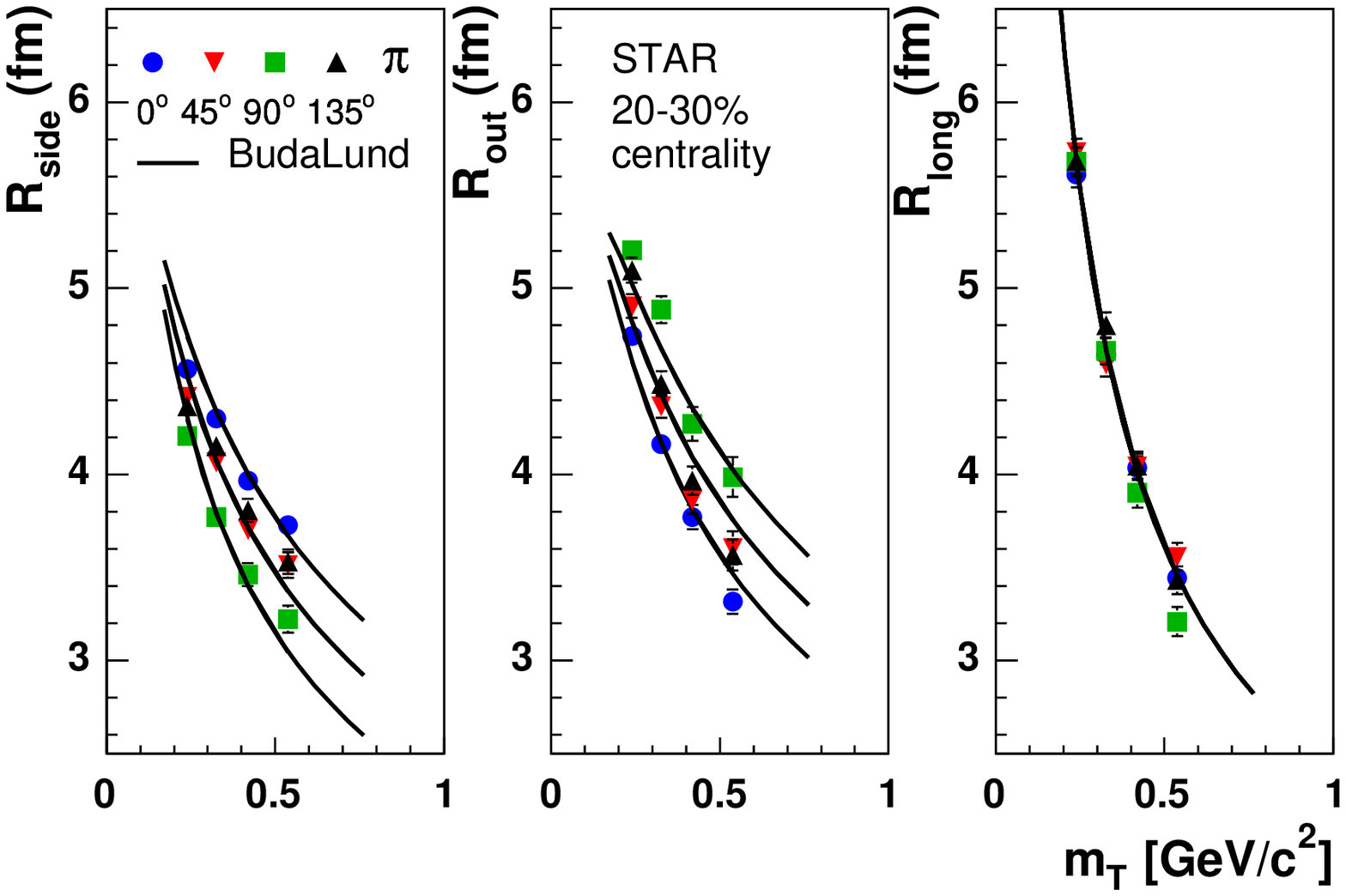}\\
		\includegraphics[width=0.90\textwidth]{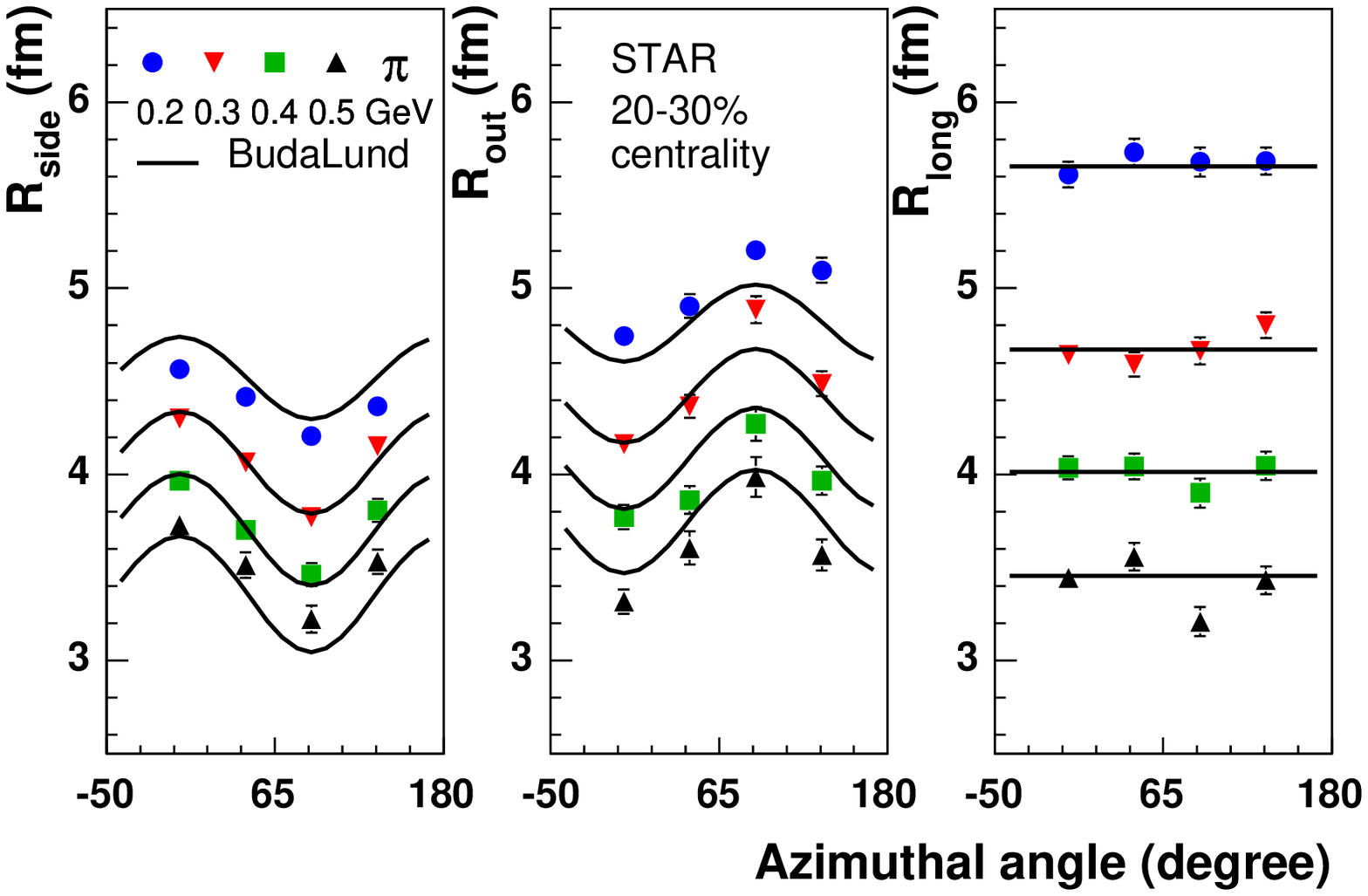}
\caption{Buda-Lund model fits to RHIC 200GeV Au+Au data of ref.~\cite{Adams:2003ra} on 
HBT radii as functions of transverse mass, for different
azimuthal angles (top). The same data and the same fit is also shown as a function
of the azimuthal angle for different values of the transverse mass (bottom). }
	\label{f:fig3}
\end{figure}
%%%%%%%%%%%%%%%%%%%%%%%%%%%%%%%%%%%%%%%%%%%%%%%%%%%%%%%%%%%%%%%%%%%%%%%%%%%%%%%%%%%%%%%
%%%%%%%%%%%%%%%%%%%%%%%%%%%%%%%%%%%%%%%%%%%%%%%%%%%%%%%%%%%%%%%%%%%%%%%%%%%%%%%%%%%%%%%
\begin{table}
\begin{center}
\begin{tabular}{|l|l|}
\hline
\hline
Buda-Lund&  parameter description \\ 
\hline
\hline
$T_0$          & Temperature in the center at $\tau_0$ \\
$T_e$          & Temperature in the center at $\tau_0 + \Delta \tau$ \\
$\mu_B$        & Baryochemical pontential in the center  at $\tau_0$\\ \hline 
$R_x$          & Geometrical size in direction $x$ \\
$R_y$          & Geometrical size in direction $y$ \\
$R_{xs}$       & Thermal size, where $T = T_0/2 $ in direction $x$ \\
$R_{ys}$       & Thermal size, where $T = T_0/2 $ in direction $y$ \\ \hline
$H_x$          & Hubble constant in direction $x$ \\
$H_y$          & Hubble constant in direction $y$ \\\hline
$\tau_0$       & Mean freeze-out proper-time \\
$\Delta\tau$   & Distribution width in proper-time $\tau$\\
$\Delta \eta $ & Distribution width in space-time rapidity $\eta$ \\ \hline
$\mu_B$        & Baryochemical potential, \\ 
\hline
\hline
\end{tabular}
\end{center}
\caption
{
Description of the parameters of the asBuda-Lund hydro model. 
The baryochemical potential is evaluated as 
$\mu_B = 1/2 \  (\mu_{0,p}-\mu_{0,\overline{p}})$.
For details see ref.~\cite{Ster:2010ia}.}
\label{t:pars}
\end{table}

\begin{table}[b]
\begin{center}
\begin{tabular}{|l|rl|rl|}
\hline
\hline
Buda-Lund     &\multicolumn{2}{c|}{Au+Au \@ 200 GeV}&\multicolumn{2}{c|}{Au+Au \@200 GeV} \\ 
parameters    &\multicolumn{2}{c|}{central (0-30\%)}&\multicolumn{2}{c|}{non-central (20-30\%)} \\
\hline
\hline
$T_0$ [MeV]          & 196      &$\pm$ 13   & 174    &$\pm$ 6 \\
$T_e$ [MeV]          & 117      &$\pm$ 11   & 130    &$\pm$ 6 \\ 
%$\mu_B$ [MeV]       &  61      &$\pm$ 55   &  56    &$\pm$ 31 \\\hline
$\mu_B$ [MeV]        &  31      &$\pm$ 28   &  27    &$\pm$ 16 \\\hline
$R_x$ [fm]           & 13.5     &$\pm$ 1.7  & 9.5    &$\pm$ 0.5 \\
$R_y$ [fm]           & $= R_x$    &           & 7.0    &$\pm$ 0.2 \\
$R_{sx}$ [fm]        & 12.4     &$\pm$ 1.6  & 12.8   &$\pm$ 0.8 \\
$R_{sy}$ [fm]        & $= R_{sx}$ &           & 16.9   &$\pm$ 1.6 \\\hline
$H_x$                & 0.119    &$\pm$ 0.020& 0.158  &$\pm$ 0.002 \\
$H_y$                & $H_x$    &           & 0.118  &$\pm$ 0.002 \\\hline
$\tau_0$ [fm/c]      & 5.8      &$\pm$ 0.3  & 5.4    &$\pm$ 0.1 \\
$\Delta\tau$ [fm/c]  & 0.9      &$\pm$ 1.2  & 2.5    &$\pm$ 0.2 \\
$\Delta\eta$         & 3.1      &$\pm$ 0.1  & 2.5    &$\pm$ 0.3 \\
\hline
$\chi^2/$NDF  &\multicolumn{2}{c|}{114/208 \@ = 0.55} &\multicolumn{2}{c|}{269.4/152 \@ = $1.77$} \\
\hline
\hline
\end{tabular}
\end{center}
\caption
{
Source parameters from simultaneous fits to PHENIX and STAR data of
Au+Au collisions at $\sqrt{s_{NN}} = 200 $ GeV,
as given in Figs. 1-3, obtained by the Buda-Lund model.
For non-central data the value of $\chi^2$/NDF refers to fits 
with statistical errors, only.
}
\label{t:results}
\end{table}
%%%%%%%%%%%%%%%%%%%%%%%%%%%%%%%%%%%%%%%%%%%%%%%%%%%%%%%%%%%%%%%%%%%%%%%%%%%%%%%%%%%%%%%
\section{Conclusions and outlook}
\label{conc}
The ellipsoidally symmetric generalization of the Buda-Lund hydrodynamic model 
compares favourably to the identified particle spectra and to the elliptic flow 
and azimuthally sensitive Bose-Einstein correlation radii of identified pions.
From model fits to 20-30 \% central Au+Au collision data at $\sqrt{s_{NN}}= 200$ GeV 
at mid-rapidity, the source parameters characterizing
these non-central ultra-relativistic heavy ion reactions were extracted.\\

The results of our analysis indicate that the central temperature
$T_{0} = 174 \pm 6$~(stat)~MeV in the 20-30\% centrality class is somewhat lower 
than that in more central collisions, 
where an earlier analysis  found $T_{0} = 196 \pm 13$~(stat)~MeV. 
We have found that the transverse flow is stronger in the 
reaction plane than out of plane
with Hubble constants $H_x=0.158 \pm 0.002$ and $H_y=0.118 \pm 0.002$.
The almond shape of the reaction zone initially elongated 
out of plane gets slightly elongated in the direction of the impact parameter 
by the time the particle emission rate reaches its maximum. The effect is 
reflected by the geometrical radii in the two perpendicular directions at 
that time, $R_x(\mathrm{in-plane}) = 9.5 \pm 0.5$~(stat)~fm,  $R_y(\mathrm{out-plane})
= 7.0 \pm 0.2$~(stat)~fm. As far as we know, this study is the first one 
where an in-plane extended
source has been reconstructed from simultaneous and reasonably successful hydro model fits 
to identified particle spectra, elliptic flow and azimuthally sensitive 
HBT data in 200 AGeV Au+Au collisions at RHIC. It is a remarkable property  of this
hadronic final state, that the ratio of the Hubble constants is approximately the same
as the ratio of the geometric source sizes: $ H_x / H_y = R_x /R_y $ 
within the errors of the analysis.

%%%%%%%%%%%%%%%%%%%%%%%%%%%%%%%%%%%%%%%%%%%%%%%%%%%%%%%%%%%%%%%%%

\section*{Acknowledgments}
We thank the Organizers of WPCF 2010 for the invititation and local support
and for creating an inspiring and fruitful scientific athmosphere at the meeting.
Inspiring discussions with professors R. J. Glauber at Harvard University,
and Mike Lisa at WPCF 2009 and P. Chung at WPCF 2010 are also gratefully acknowledged.
T.  Cs. was also supported by a Senior Leaders and Scholars Fellowship 
of the Hungarian American Enterprise Scholarship Fund.
This work was also supported by a bilateral collaboration
between Hungary and Slovakia, project Nos.\ SK-20/2006 (HU), SK-MAD-02906 (SK).
B.T.\ acknowledges support by MSM 6840770039 and
LC 07048 (Czech Republic), as well as VEGA1/0261/11 (Slovakia). 
M.Cs.,\ T.Cs.\ and A.S.\ gratefully acknowledge the support of
the Hungarian OTKA  grant NK 73143. 
%%%%%%%%%%%%%%%%%%%%%%%%%%%%%%%%%%%%%%%%%%%%%%%%%%%%%%%%%%%%%%%%%%

%\bibliographystyle{prlsty}

\end{document}